\documentclass[conference]{IEEEtran}
\IEEEoverridecommandlockouts
\usepackage{cite}
\usepackage{amsmath,amssymb,amsfonts}
\usepackage{algorithmic}
\usepackage{graphicx}
\usepackage{textcomp}
\usepackage{xcolor}
\usepackage{siunitx}
\usepackage{subcaption, standalone} 
\usepackage{tikz-qtree}
\usepackage{tikz}
\usepackage{siunitx}
\usepackage{enumitem}
\usepackage{multirow}
\usepackage{adjustbox}
\def\BibTeX{{\rm B\kern-.05em{\sc i\kern-.025em b}\kern-.08em
    T\kern-.1667em\lower.7ex\hbox{E}\kern-.125emX}}
\begin{document}

\title{Whisker-Like Sensors: Effective Design}

\author{
\IEEEauthorblockN{Prasanna Kumar Routray\IEEEauthorrefmark{1}}
\IEEEauthorblockA{Applied Mechanics and Biomedical Engineering\\
Indian Institute of Technology Madras\\
Chennai, India, 600036\\
Email: prasanna.routray97@gmail.com}\\   
\IEEEauthorblockN{Pauline Pounds}
\IEEEauthorblockA{School of Information Technology and Electrical Engineering\\
The University of Queensland\\
Brisbane, Australia, 4072\\
Email: pauline.pounds@uq.edu.au}
\and
\IEEEauthorblockN{Debadutta Subudhi}
\IEEEauthorblockA{Applied Mechanics and Biomedical Engineering\\
Indian Institute of Technology Madras\\
Chennai, India, 600036\\
Email: dev.subudhi49@gmail.com}\\  
\IEEEauthorblockN{Manivannan M.}
\IEEEauthorblockA{Applied Mechanics and Biomedical Engineering\\
Indian Institute of Technology Madras\\
Chennai, India, 600036\\
Email: mani@iitm.ac.in}
}
\maketitle

\begin{abstract}
Bio-inspired whisker sensors are employed in diverse applications such as fluid-flow sensing, texture analysis, and environmental exploration. However, existing designs often face challenges related to durability, fabrication complexity, and response consistency. To address these issues, we propose a modular architecture that decomposes whisker sensors into five functional components: the whisker element (WE), compliant element (CE), sensing element (SE), support structure (SS), and data acquisition module (DAQ). We develop and compare four in-house sensor designs built using this architecture, each differing in material choice, sensing modality, and mechanical structure. To unify heterogeneous sensor outputs, we introduce a calibration strategy that maps raw sensor readings—whether from pressure, magnetic flux, or visual features—into a common representation: the bending moment at the whisker base. This representation supports consistent interpretation and comparison across sensing techniques. We adopt texture roughness analysis as a representative sensing task to evaluate design trade-offs. Each whisker sensor’s frequency-domain response is benchmarked against a high-resolution laser sensor using standardized roughness specimens. Empirical results show that rigid whiskers improve accuracy in texture classification, while flexible whiskers provide robustness for exploratory robotics tasks. Among the evaluated designs, the Hall-effect sensor with a rubber CE demonstrates the most favorable balance of durability, reconfigurability, and fabrication simplicity.
\end{abstract}


\section{Introduction}
Whiskers in animals such as rodents, felids, and pinnipeds play a vital role in tactile sensing tasks, including prey detection, environmental mapping, and fluid-flow sensing. This natural sensing capability has inspired bio-inspired whisker sensor designs. While many demonstrate functional promise, few can be reliably reproduced due to fabrication complexity, inconsistent materials, or fragile transduction methods. Designs range from Russell’s early implementation \cite{russell1984closing} to recent MEMS-based approaches \cite{wei2019development}, yet no modular, scalable framework exists. An ideal whisker sensor should be affordable, durable, easy to fabricate, and reconfigurable, all while maintaining high sensing fidelity. These limitations highlight the need for a practical design approach that overcomes current barriers in whisker sensor development.

Whisker sensors have been extensively explored in tactile robotic applications, including contact detection, shape recognition, and terrain interaction \cite{lepora2018tacWhiskers, deer2019lightweight, kim2019magnetically}. Designs span a range of transduction mechanisms, such as PVDF-based sensors \cite{tiwana2016bio}, capacitive flow sensors \cite{wissman2019capacitive}, piezoelectric sensor-actuator hybrids \cite{ju2014bioinspired}, and strain-gauge-based configurations \cite{solomon2006robotic}. While several reviews have summarized these technologies and their use cases \cite{prescott2009whisking, sayegh2022review, wang2023potential, yu2024whisker}, few address key challenges such as design reproducibility, material selection, and integration complexity—factors critical to real-world deployment. Overcoming these barriers is essential to transitioning whisker sensor technology from research prototypes to scalable systems.

This work addresses the challenge of design complexity by evaluating four in-house fabricated whisker sensor prototypes. These designs vary in sensing modality, whisker material, compliant elements, and structural configuration. We assess them using a consistent framework that includes mechanical structure, ease of fabrication, data acquisition method, and sampling rate. Texture analysis, similar to the work in \cite{routray2022towards}, is chosen as a representative task because it is relevant to tactile sensing and reveals differences in sensor sensitivity and resolution. Our objective is to identify sensor designs that are robust, reconfigurable, and simple to fabricate, making them suitable for practical use with readily available components.

We present a comparative study of four whisker-inspired sensor designs that differ in transduction mechanisms, compliant elements, and structural complexity. We introduce a unified framework for interpreting sensor responses via bending moment analysis and validate performance through texture classification experiments on machined surfaces. Our key contributions include: (1) a generalizable comparison methodology for modular whisker sensors, (2) practical design recommendations for robust and reproducible fabrication, and (3) empirical evidence linking sensor configuration to task-specific performance.

\begin{figure*}[htb!]
\centering
\includegraphics[width=1.0\textwidth, height=11cm]{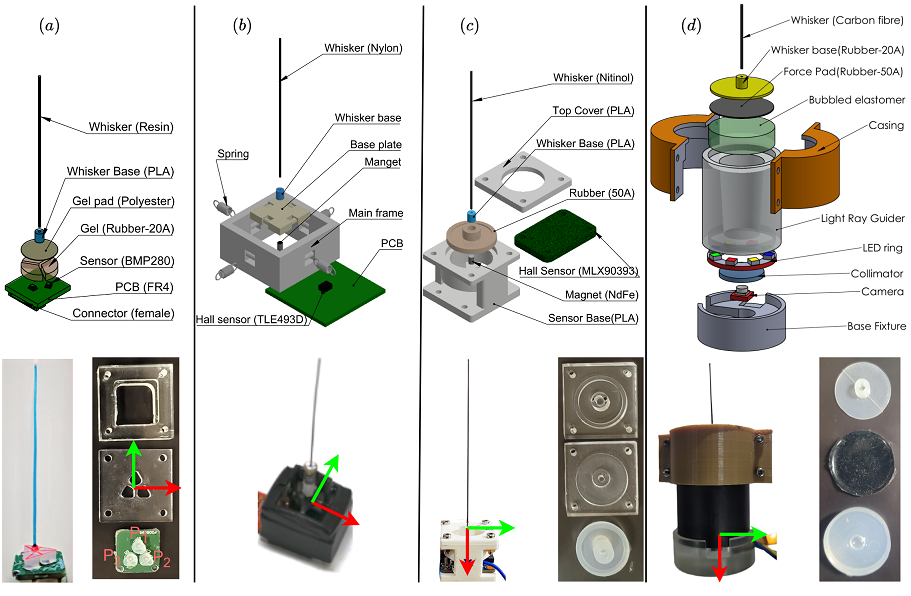}
\caption{\footnotesize{Sensor design concepts (top row) and corresponding fabricated sensors (bottom row) are shown. 
        (a) \textbf{M-Whisker:} This design uses a BMP-280 pressure sensor covered with a 3\,mm thick silicone rubber gel pad. The setup is 15\,mm wide. Rubber compliance and molding elements are also shown. 
        (b) \textbf{H-Whisker (Spring):} This configuration employs a TLE-493D Hall sensor connected to a nylon whisker and a helical spring compliant element. The setup is 35\,mm wide, with spring dimensions of $0.3 \times 3 \times 7$\,mm. 
        (c) \textbf{H-Whisker (Rubber):} This design uses an MLX90393 Hall sensor coupled with a Nitinol whisker and a rubber-based compliant element. The setup is 25\,mm wide. Replaceable rubber diaphragm and mold components are also shown. 
        (d) \textbf{B-Whisker:} This design integrates a spy camera as a vision module, along with a single air bubble and a carbon fiber whisker. The setup has a diameter of 40\,mm, with compliant elements placed adjacent.
    }}
\label{fig:Schematics}
\end{figure*}

\section{Design of Modular Whisker Sensors}\label{sec:sensorFabrication}
This section introduces a generalized modular framework for bio-inspired whisker sensors and describes four in-house designs developed using this architecture. By examining the core components and their roles, we enable a systematic comparison of sensor behavior and buildability.

Bio-inspired whisker sensors are typically composed of five functional components: (1) a whisker element (WE), which acts as a cantilevered probe; (2) a compliant element (CE), which restores the whisker to its resting state and modulates sensing dynamics; (3) a sensing element (SE) positioned at the whisker base to detect motion or deflection; (4) a support structure (SS) to house and align components; and (5) a data acquisition or communication module that transmits signals to a microcontroller (MCU) or processing board. While these elements are common across designs, choices in material, transduction, and fabrication significantly affect sensor performance and durability.

Fig.~\ref{fig:Schematics} shows the configuration of the four in-house designs used in this study. Each employs a distinct combination of the above components. Despite differences in transduction mechanisms and structural layout, the sensor’s output at the base should follow a consistent dynamic response. This observation, consistent with LaValle’s framework for comparable sensor models \cite{lavalle2012sensing}, supports the potential for standardizing performance metrics across designs through bending moment analysis that we propose.

\subsubsection{Whisker Element (\textbf{WE})}
The whisker element is the physical probe that interacts with the environment, analogous to animal vibrissae. Typically shaped as a straight, curved, or tapered cantilever, the WE is not itself an active sensor in most designs, except for specialized implementations like PVDF-based distributed sensors \cite{tiwana2016bio}. In our prototypes, we used Nitinol, carbon fiber, nylon, and 3D-printed resin—materials selected for their mechanical resilience, shape memory, or ease of prototyping. Tapered and curved shapes offer better compliance and contact sensitivity but are harder to fabricate. We successfully 3D-printed a tapered WE using UV-cured resin (Fig.~\ref{fig:Schematics}(a)), while attempts with PLA were unsuccessful. Off-the-shelf nylon, nitinol, and carbon fiber whiskers were used in the designs shown in Fig.~\ref{fig:Schematics}(b--d).

\subsubsection{Compliant Element (\textbf{CE})}
The compliant element (CE) governs how forces are transferred to the sensing element and influences the return-to-rest dynamics after deflection. Common implementations include mechanical springs (helical or serpentine) and elastomeric materials such as silicone rubber \cite{kim2019magnetically}. In our designs, rubber CEs were favored due to their symmetric deformation profile, easy moldability, and tunability. For instance, in Fig.~\ref{fig:Schematics}(c), we used a replaceable rubber diaphragm of 1.5\,mm thickness to allow for adjustable stiffness and damping. Compared to helical springs (Fig.~\ref{fig:Schematics}(b)), rubber offered improved symmetry and ease of integration, though issues such as trapped air during molding remain a practical challenge. In magnetic designs, the CE must also support the magnet’s weight without compromising the fidelity of the response.

\subsubsection{Sensing Element (\textbf{SE})}
The sensing element (SE) captures mechanical inputs and converts them into measurable outputs. It is either analog (requiring signal conditioning) or digital (e.g., MEMS-based). Designs fall into three broad categories: (1) discrete sensors with separate signal conditioning \cite{russell1984closing, wissman2019capacitive, tiwana2016bio}, (2) integrated MEMS sensors \cite{deer2019lightweight, kim2019magnetically}, and (3) vision-based systems \cite{lepora2018tacWhiskers}. Our designs incorporate MEMS piezoresistive pressure sensors (Fig.~\ref{fig:Schematics}(a)), Hall-effect sensors (Fig.~\ref{fig:Schematics}(b--c)), and a camera module for the vision-based sensor (Fig.~\ref{fig:Schematics}(d)). While MEMS sensors are compact, they require careful assembly. In contrast, Hall-effect sensors offer robust, non-contact measurement. Vision-based SEs allow rich spatial data but demand complex image processing.

\subsubsection{Support Structure (\textbf{SS})}
The support structure anchors the CE and SE while preserving geometric alignment. We explored three fabrication methods: PCB-mounted (Fig.~\ref{fig:Schematics}(a)), 3D-printed ABS (Fig.~\ref{fig:Schematics}(b)), and acrylic cutouts (Fig.~\ref{fig:Schematics}(d)). 3D printing offers rapid prototyping and design freedom, while PCB-mounted structures provide improved rigidity and electrical integration. Vision-based designs benefit from transparent or rigid structures to ensure consistent optical alignment. The choice of support material and fabrication method affects not only mechanical stability but also assembly complexity and sensor reproducibility.

\subsubsection{Data Acquisition (\textbf{DAQ})}
Sensor outputs are digitized and transmitted via standard communication protocols. MEMS sensors simplify hardware integration by embedding signal conditioning. In our designs, we use SPI for the M-Whisker (Fig.~\ref{fig:Schematics}(a)), I2C for H-Whiskers (Figs.~\ref{fig:Schematics}(b) \& \ref{fig:Schematics}(c)), and USB for the vision-based B-Whisker (Fig.~\ref{fig:Schematics}(d)). SPI offers high-speed, full-duplex communication with fewer address limitations than I2C. However, I2C is simpler to implement for low-bandwidth applications. USB, while versatile, introduces latency and higher power demands. Selection of DAQ protocol thus reflects trade-offs between speed, scalability, and hardware simplicity.

\subsubsection{Fabrication Challenges and Trade-offs}
Each sensor design poses unique fabrication challenges. For instance, the B-Whisker (Fig.~\ref{fig:Schematics}(d)) requires careful bubble insertion and optical alignment, while the M-Whisker (Fig.~\ref{fig:Schematics}(a)) involves handling delicate pressure sensors during rubber curing. The use of superglue to join CE, WE, and SE components raises issues of alignment accuracy and repeatability. Laser-cut serpentine springs, although precise, require specialized tools, and sourcing miniature off-the-shelf springs remains a challenge. Achieving structural symmetry and avoiding tilt during assembly are especially important for consistent sensing performance.

Table~\ref{tab:sensor_comparison} compares four in-house whisker sensor prototypes, each designed using the modular architecture described in this section. These designs differ in sensing modality, mechanical structure, and material choices, allowing us to evaluate trade-offs in durability, complexity, data quality, and ease of fabrication. All designs are scalable and reconfigurable, but their performance characteristics vary depending on the application context.

For reference, Sensor~1 corresponds to the M-Whisker (Fig.~\ref{fig:Schematics}(a)), Sensor~2 to the H-Whisker (Spring) (Fig.~\ref{fig:Schematics}(b)), Sensor~3 to the H-Whisker (Rubber) (Fig.~\ref{fig:Schematics}(c)), and Sensor~4 to the B-Whisker (Fig.~\ref{fig:Schematics}(d)).

\begin{table}[htb!]
\caption{\footnotesize{Comparison of key characteristics of the four in-house whisker sensor designs. WE: Whisker Element, CE: Compliant Element. The empirical and theoretical findings reported here are from observations of three or more prototypes for each design.}}
\label{tab:sensor_comparison}
\centering
\resizebox{\linewidth}{!}{
\begin{IEEEeqnarraybox}[\IEEEeqnarraystrutmode\IEEEeqnarraystrutsizeadd{2pt}{1pt}][b][\columnwidth]{s+t+t+t+t}
\IEEEeqnarraydblrulerowcut\\
\IEEEeqnarrayseprow[2pt]{}\\
\textbf{Parameter} & \textbf{Sensor 1} & \textbf{Sensor 2} & \textbf{Sensor 3} & \textbf{Sensor 4} \\
\IEEEeqnarrayseprow[2pt]{}\\
\IEEEeqnarrayrulerow\\
\IEEEeqnarrayseprow[2pt]{}\\
WE Material & Resin/Carbon Fiber & Nylon & Nitinol & Carbon Fiber \\
WE Type & Flexible/Rigid & Flexible & Flexible & Rigid \\
CE Material & Rubber & Spring & Rubber & Rubber \\
CE Tunable & Yes & Yes & Yes & Yes \\
CE Build Effort & Medium & Low & Low & High \\
Durability & Low & Medium & High & Medium \\
Build Complexity & Medium & Low & Low & High \\
Size & Small & Medium & Small & Medium \\
Scalability & Yes & Yes & Yes & Yes \\
Sampling Rate (Hz) & 166 & 8400 & 500 & 90 \\
Protocol & SPI & I2C & I2C/SPI & USB \\
MEMS-Based & Yes & Yes & Yes & No \\
\IEEEeqnarrayseprow[4pt]{}\\
\IEEEeqnarraydblrulerowcut
\end{IEEEeqnarraybox}
}
\end{table}

\vspace{0.5em}
The five modular components—WE, CE, SE, SS, and DAQ—form the basis of the four whisker sensor configurations evaluated in this study. By varying materials and sensing methods, we can systematically explore trade-offs between fabrication effort, mechanical performance, and sensing fidelity. This modular approach supports reconfigurability and enables direct comparison across transduction methods using bending moment at the base—essential for identifying optimal designs for tasks such as texture analysis or environment exploration. Both empirical and theoretical findings presented here are derived from three or more prototypes for each design.

\section{Sensor Response Interpretation and Calibration}
\subsection{Base Moment Modeling}\label{sec:response}
To compare heterogeneous whisker sensor designs, a unified output domain is necessary. Since sensor outputs vary based on transduction mechanisms, we translate all responses into an equivalent bending moment at the whisker base. This approach follows the sensor model generalization proposed by LaValle \cite{lavalle2012sensing}, which allows meaningful comparison or simulation of sensors whose input-output mappings can be converted through a common representation.

\textit{Coordinate Convention:} In this study, the whisker is assumed to be cantilevered along the \textbf{z-axis}, a common setup in vertically mounted designs. During interaction, the whisker may deflect in the ZY or ZX planes, corresponding to lateral displacements along the y- or x-axes, respectively. These deflections generate bending moments about the x-axis (\( M_X \)) and y-axis (\( M_Y \)) at the base. Torsional deformation is not modeled, so \( M_Z = 0 \). The base moment vector is defined as
\[
\vec{M}(0) = [M_X(0), M_Y(0), 0]
\]
Throughout this work, a consistent color convention is adopted in figures: the X-, Y-, and Z-axes are represented by red, green, and blue, respectively (i.e., RGB $\rightarrow$ XYZ). This aids interpretation of whisker geometry, deflection, and calibration setup.

We define the following mappings for each sensor type, translating its raw measurements into base moments:

\begin{itemize}[leftmargin=*]

\item \textbf{M-Whisker:}  
This sensor integrates three MEMS pressure sensors arranged radially at \SI{120}{\degree} intervals around the compliant element, starting with $P_1$ in the first quadrant, as shown in Fig.~\ref{fig:Schematics}(a). The pressure readings are projected into Cartesian coordinates $(P_X, P_Y, P_Z)$ to estimate the resultant bending moment $(M_X, M_Y)$ using the following formula:
\begin{equation}
\begin{split}
    P_X &= (P_1 - P_2) \cos(30^\circ) \\
    P_Y &= -\big((P_1 + P_2) \cos(60^\circ) - P_3\big) \\
    P_Z &= -\frac{1}{3}(P_1 + P_2 + P_3) \\
    f_1 &: (P_X, P_Y, P_Z) \rightarrow (M_X, M_Y),
\end{split}
\end{equation}
where $f_1$ is the transformation from effective pressure space to bending moment space.

\item \textbf{H-Whisker (Spring/Rubber):}  
The Hall-effect sensor measures magnetic flux density components $(B_X, B_Y, B_Z)$ in \SI{}{\micro\tesla}, which are influenced by whisker deflection. A calibration matrix then maps these components to the moment vector at the base. This approach follows prior work on spring- and rubber-based whiskers by Kim et al. \cite{kim2019magnetically} and Lin et al. \cite{lin2022whisker} as
\begin{equation}
f_2: (B_X, B_Y, B_Z) \rightarrow (M_X, M_Y).
\end{equation}

\item \textbf{B-Whisker (Vision-based):}  
This sensor uses a miniature camera to capture the deformation of an internal air bubble caused by whisker deflection. The camera produces grayscale image frames \( g_1(i, j) \), where \( (i,j) \) are pixel indices. From each image, a visual feature vector \( F \) is extracted—such as centroid displacement, ellipse orientation, or optical flow—through a mapping \( g_2 \). These features are then mapped to mechanical bending moments at the whisker base using a learned function \( f_3 \). The overall transformation is:

\begin{equation}
f_3 \circ g_2: g_1(i, j) \rightarrow (M_X, M_Y)
\end{equation}
where \( g_2 \) captures geometric features from the image, and \( f_3 \) maps these features to the estimated bending moments. Although the current implementation emphasizes \( M_X \), the mapping framework is generalizable to produce both moment components through improved feature design and spatial calibration.
\end{itemize}

\subsection{Calibration}
To establish a consistent mapping from sensor measurements to bending moments, we perform calibration using the numerical model presented by Quist et al. \cite{quist2012mechanical}. The goal is to derive an empirical function that translates magnetic flux density readings from the Hall-effect sensor into the bending moment at the whisker base. This method can be generalized to other sensors.

\textbf{1. Calibration Setup:}  
We deflect the whisker to known spatial coordinates $(p_{z_i}, p_{y_i})$ in the sensor reference frame. At each point, we record both the flux density $B_{Y_i}$ and the expected bending moment $M_{X_i}(0)$ based on cantilever theory.  
The deflection points lie in the ZY plane, causing bending about the X-axis and resulting in non-zero \( M_X \).

\textbf{2. Model-Based Reference:}  
A numerical whisker model adapted from Quist et al. \cite{quist2012mechanical} simulates the moment response under the same boundary and loading conditions. It operates in 'point mode' and matches the mechanical parameters of the physical sensor described in Section~\ref{sec:sensorFabrication}.

\textbf{3. Boundary Conditions:}  
To improve accuracy, an elastic (non-rigid) boundary condition is applied at the clamped end of the simulated cantilever to mimic the compliance observed in physical builds.

\textbf{4. Data Collection and Mapping:}  
Using $n = 1000$ deflection samples, we construct a functional mapping:
\begin{equation}
M_X(0) = \hat{f}_{2}(B_Y),
\end{equation}
where \( \hat{f}_2 \) represents the fitted function derived from calibration for bending in the ZY plane only (no bending in ZX plane).

\begin{figure}[!ht]
\centering
\subfloat[]{\includegraphics[width=0.4\textwidth, height=3cm]{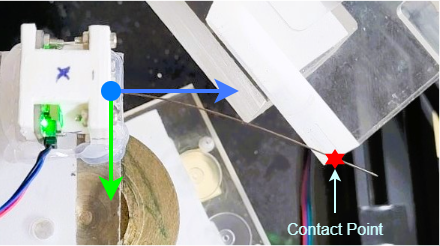}}\\
\subfloat[]{\includegraphics[width=0.24\textwidth, height=4cm]{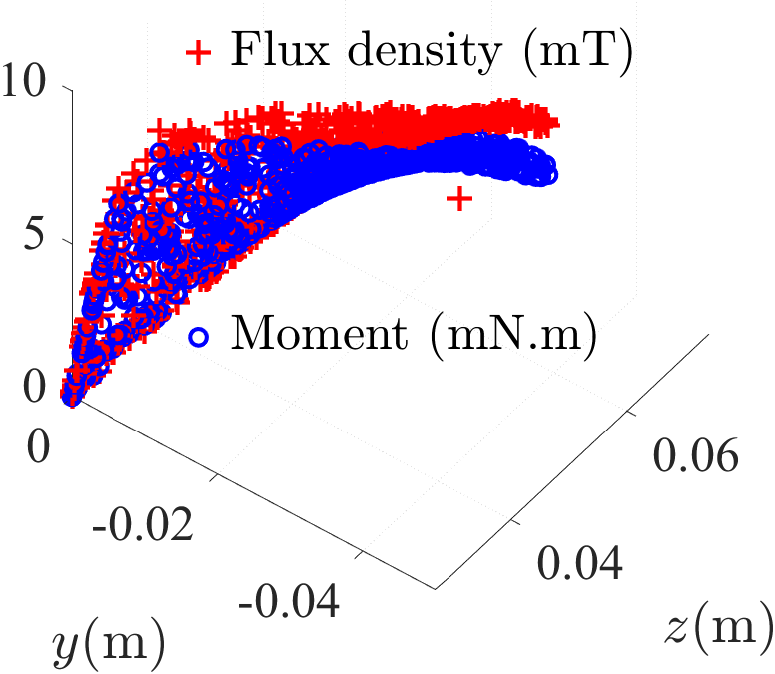}}
\hfil
\subfloat[]{\includegraphics[width=0.24\textwidth, height=4cm]{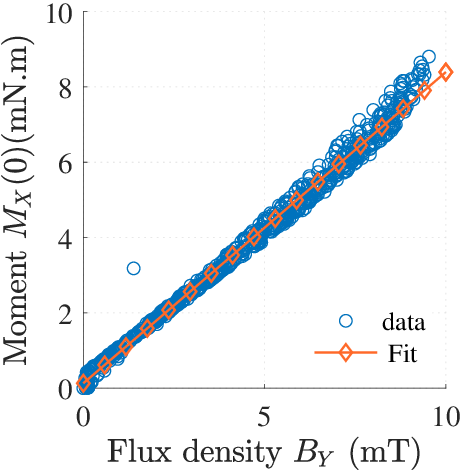}}
\caption{\footnotesize{Sensor calibration steps: (a) Experimental setup for mapping magnetic flux density (measured by the physical sensor) to bending moment (simulated by the numerical model) using a common set of contact points. (b) Measured magnetic flux density and simulated bending moment plotted across contact locations. (c) Calibration curve mapping magnetic flux density to base bending moment.}}
\label{fig:sensorCalibration}
\end{figure}

This calibration process establishes a method to convert sensor outputs into a unified bending moment representation, independent of mechanical structure or transduction mode. Results shown in Fig.~\ref{fig:sensorCalibration} demonstrate the mapping and applicability for interpreting sensor responses across designs.

While our modeling framework supports both \( M_X \) and \( M_Y \), the current calibration setup applies deflection primarily in the ZY plane, corresponding to bending about the X-axis. Therefore, only \( M_X \) is derived from the calibration dataset in this study. Mapping to \( M_Y \) follows the same process but requires deflections in the ZX plane and appropriate flux components.

However, the whisker element (WE) plays a critical role in determining the uniqueness of this mapping. For highly flexible WEs, the sensor exhibits a one-to-many mapping, where a single flux measurement corresponds to multiple potential base moments due to mechanical nonlinearity and hysteresis. This underscores the importance of considering WE stiffness and geometry when designing sensors for consistent and invertible response interpretation.

\section{Experimental Setup \& Results}
This section presents a comparative evaluation of three in-house whisker sensor designs using a standard surface texture analysis task. While the unified calibration framework introduced earlier translates sensor outputs into base bending moments, this experiment focuses on task-specific signal interpretation in the frequency domain. This approach offers a low-complexity yet effective method for assessing sensor performance against a laser-based ground truth.

Before the experimental evaluation, it is important to note that calibrated responses were not used here because not all designed sensors were calibrated. To reduce the time-consuming calibration process, we employ texture analysis as a task-specific method to evaluate sensor performance. In robotics applications, mapping all sensor responses to a unified domain—preferably the bending moment via calibration—is desirable for effective benchmarking and performance comparison.

\subsection{Experimental Setup}
\begin{figure}[ht]
\centering
\includegraphics[width=0.48\textwidth]{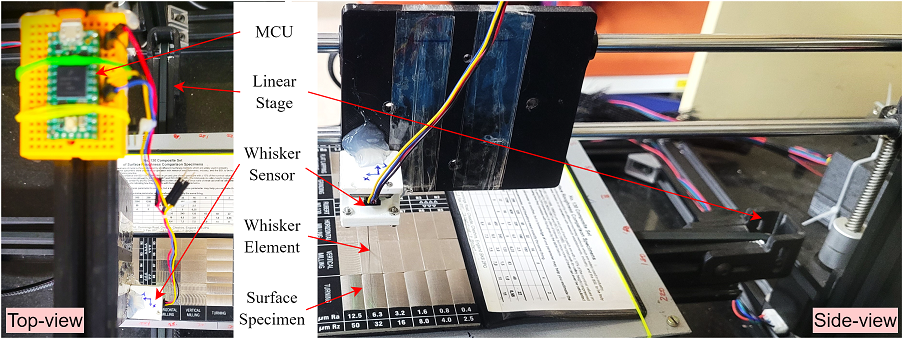}
\caption{\footnotesize{Experimental setup showing top and side views: linear stage, mounted whisker sensor, and textured specimen.}}
\label{fig:setup}
\end{figure}

The experimental platform (Fig.~\ref{fig:setup}) comprises a precision linear stage, machined surface specimens, and a mounted whisker sensor. The stage enables controlled lateral brushing of the whisker across each specimen with a positional resolution of \SI{10}{\micro\meter}, ensuring precise and repeatable scanning.

A high-accuracy Optical NCDT laser sensor by Micro-Epsilon serves as the gold standard. Operating at \SI{50}{\kilo\hertz} with a \SI{670}{nm} wavelength, it can detect peak-to-valley variations as small as \SI{0.3}{\micro\meter}, making it a suitable sensor for surface texture profiling.

The whisker scans were performed at a constant speed of \SI{50}{\milli\meter\per\minute}. To satisfy the Nyquist criterion and prevent aliasing, the displacement per sample $\mathcal{D}$ is constrained by:
\[
\mathcal{D} < \frac{1}{2} d_{sep}
\]
where \(d_{sep}\) is the characteristic peak spacing of the textured surface.

The surface roughness specimens used in this study were acquired from RUBERT \& Co. LTD., England, featuring standardized textures with peak-to-valley heights ranging from \SI{2.5}{\micro\meter} to \SI{50}{\micro\meter}.

\subsection{Texture Response Evaluation}
Three sensor configurations—M-Whisker, H-Whisker (Rubber), and B-Whisker—were tested across machined surfaces. Each surface was scanned five times by each sensor and the laser. The signal from each sweep was analyzed in the frequency domain using power spectral density analysis. The dominant frequency component, corresponding to the periodic surface pattern, was extracted.

Table~\ref{table_result} reports the \textbf{mean frequency peak} from five trials per sensor and surface. The \textbf{percentage error} $\varepsilon$ is computed with respect to the laser’s average. This method reduces scan-to-scan noise while avoiding the limitations of standard deviation in spectral analysis.
\begin{table}[htb!]
\caption{\footnotesize{Results from texture analysis using in-house whisker sensors and a laser sensor. HM: Horizontal Milling, VM: Vertical Milling, T: Turning. $\varepsilon$ indicates percentage error.}}
\label{table_result}
\centering
\begin{IEEEeqnarraybox}[\IEEEeqnarraystrutmode\IEEEeqnarraystrutsizeadd{2pt}{1pt}][b][\columnwidth]{s+t+t+t+t}
\IEEEeqnarraydblrulerowcut\\
\IEEEeqnarrayseprow[2pt]{}\\
\textbf{Surface} & \textbf{Laser (\SI{}{\hertz})} & \textbf{M-Whisker} ($\varepsilon$\%) & \textbf{H-Whisker} ($\varepsilon$\%) & \textbf{B-Whisker} ($\varepsilon$\%) \\
\IEEEeqnarrayseprow[2pt]{}\\
\IEEEeqnarrayrulerow\\
\IEEEeqnarrayseprow[2pt]{}\\
\SI{32}{\micro\meter} HM & 0.34 & 0.39 (14.7) & 0.10 (70.5) & 0.17 (50.1) \\
\SI{32}{\micro\meter} VM & 1.22 & 1.25 (2.46) & 1.24 (1.64) & 1.52 (24.6) \\
\SI{32}{\micro\meter} T  & 1.71 & 1.83 (7.01) & 1.72 (0.58) & 1.75 (2.34) \\
\IEEEeqnarrayrulerow\\
\SI{50}{\micro\meter} HM & 0.19 & 0.18 (5.26) & 0.21 (10.52) & 0.17 (10.52) \\
\SI{50}{\micro\meter} VM & 0.68 & 0.70 (2.94) & 0.69 (1.47) & 0.67 (1.47) \\
\SI{50}{\micro\meter} T  & 1.14 & 1.13 (0.87) & 1.13 (0.87) & 1.142 (0.17) \\
\IEEEeqnarrayseprow[4pt]{}\\
\IEEEeqnarraydblrulerowcut
\end{IEEEeqnarraybox}
\end{table}

For simplicity and task relevance, this evaluation uses raw sensor outputs rather than calibrated bending moment mappings. The calibration-to-moment framework remains applicable for more advanced 3D sensing or force inference.

\subsection{Performance Insights}
Results show all sensors perform well on Turning (T-type) surfaces, with errors below 2\%. On coarser surfaces (HM-type), the M-Whisker exhibits the lowest error, attributable to its higher flexural rigidity, which preserves sensitivity to high-frequency surface features. The B-Whisker, despite having a rigid carbon fiber whisker, suffers from a hyperelastic compliant element that attenuates signal fidelity due to low restoring stiffness.

The H-Whisker with rubber CE provides balanced performance on VM-type surfaces and low error on smoother profiles, indicating adaptability in mid-range texture sensing tasks. The H-Whisker with a spring demands precise assembly and accurate spring anchoring that influences performance consistency. In this study, it was excluded from testing as it often showed unreliable recovery to the initial state after each sweep.

Overall, while the calibration-to-moment methodology remains broadly applicable, direct signal analysis provides a pragmatic approach for evaluating whisker sensor designs in texture classification. This task-specific benchmarking complements our modeling pipeline and highlights compliance design and whisker stiffness as critical determinants of sensing performance.

\section{Conclusion}
This study presents a comparative evaluation of whisker sensor architectures, focusing on design complexity, fabrication feasibility, and task-specific sensing performance. Among the tested configurations, the Hall-effect sensor paired with a rubber-compliant element stands out for its durability, ease of fabrication, affordability, and modular reconfigurability. In contrast, pressure sensor-based designs pose significant fabrication challenges, including whisker attachment, removal of protective metal caps, and delicate handling required after silicone curing, where mishandling risks sensor damage. Similarly, the bubble-based vision sensor faces consistency and practicality issues due to the difficulty of reliably inserting and aligning the bubble. The spring-based sensor, while mechanically functional, demands careful assembly to preserve spring properties and requires access to suitable off-the-shelf components. Overall, the rubber-compliant Hall sensor design emerges as symmetric, durable, and more fabrication-friendly than the alternatives.

To evaluate sensing performance, the three whisker sensor designs were tested using standardized surface roughness specimens from RUBERT \& Co. LTD., England. Texture analysis compared sensor signals against a high-precision optical laser sensor (Micro-Epsilon NCDT) capable of \SI{0.03}{\micro\meter} peak-to-valley resolution at \SI{50}{\kilo\hertz}. While the proposed calibration-to-moment pipeline enables general comparisons across heterogeneous sensors, direct frequency-domain analysis of raw sensor signals proved more efficient for textural roughness analysis task. empirical results show that rigid whiskers are crucial for accurate texture or flow sensing, whereas flexible whiskers better suit environment exploration due to their resilience in unstructured interactions. As researchers explore a wide range of design options, this work provides a clear, comparative roadmap for developing whisker sensors that are effective, task-tuned, robust, and simple to build.

\section*{Acknowledgments}
We gratefully acknowledge Prof. Mitra J. Hartmann for making the \textit{\textbf{elastica2D}} model code available, which was used in this study.

\normalsize

\bibliographystyle{IEEEtran}
\bibliography{refs}
\end{document}